\documentclass[a4paper,11pt]{article}
\usepackage[margin=1in]{geometry}

\usepackage{layout}
\usepackage[english]{babel}
\usepackage[utf8]{inputenc}

\usepackage{algorithm}
\usepackage{algorithmicx}

\usepackage{algpseudocode}
\usepackage{lipsum}% http://ctan.org/pkg/lipsum
\usepackage{float}% http://ctan.org/pkg/float
\floatstyle{boxed} % Box...
\restylefloat{figure}% ...figure environment contents.
\usepackage{graphicx} % if you want to include jpeg or pdf pictures

\graphicspath{ {Reportimages/} }

\usepackage{ragged2e}
\usepackage{multicol}
\usepackage{color,graphicx}
\usepackage{amsmath}
\usepackage{epstopdf}
\usepackage{epsfig}
\usepackage{float}
\usepackage{cite}

\title{Multi-threaded Graph Coloring Algorithm for Shared Memory Architecture}

\begin{document}
\begin{center}
\large{\textbf{Multi-threaded Graph Coloring Algorithm for Shared Memory Architecture}}\\
\vspace{3mm}
Nandini Singhal, Sathya Peri, Subrahmanyam Kalyanasundaram\\
Department of Computer Science \& Engineering\\
Indian Institute of Technology, Hyderabad\\
Email: [cs15mtech01004, sathya\_p, subruk]@iith.ac.in

\vspace{5mm}
\end{center}
%\begin{multicols}{2}
%\vspace{0.01mm}
%\justify

\section{Introduction}
The Graph Coloring Problem deals with assigning colors to the vertices of a simple graph such that no two adjacent vertices get the same color (also termed as vertex coloring). The graph coloring problem has a central role in computer science. It models many significant real-world problems, or arises as part of a solution for other important problems such as register allocation, scheduling, frequency assignment, data mining and many others. Coloring an arbitrary graph with the minimum number of colors is known to be an NP-hard problem, thus the primary goal in doing so is to minimise the number of colors used.\\
\indent With the domination of multi-core systems, hardware capability can be completely exploited with parallel algorithms. Each core can independently process a subtask and this can speed up the overall performance of the algorithm. However, sequential algorithms act only on single core, albeit availability of multi-cores. In this particular problem, the two key parameters to be kept in mind are: to reduce the coloring time along with minimising the number of colors used (ensuring proper coloring).\\
\indent Previous work on developing such algorithms has been performed on distributed memory computers using explicit message-passing \cite{boman2005scalable}. The same notion has been extended to the shared memory architecture. The main justification for using distributed algorithms has been access to more memory and thus, the ability to solve problems with very large graphs. It is to be noted that the current availability of multi-core computers where the entire memory can be accessed by any processor makes this argument less significant. The development of multi-core systems has been accompanied by the emergence of new shared memory programming paradigms like OpenMP and Pthreads, which are very easier to program. Posix threads provide low-level API for working with threads. Hence we simulate using Pthreads for fine grained control over thread management.\\
\indent In this paper, we present parallel algorithms for graph coloring suitable to shared memory programming model. The first algorithm deals with the most widely studied technique of coloring using barrier synchronization \cite{patwary2011new,DBLP:journals/pc/CatalyurekFGHP12,boman2005scalable, DBLP:journals/concurrency/GebremedhinM00}. A barrier for a group of threads or processes in the source code means any thread/process must stop at that point and cannot proceed until all other threads/processes have reached the barrier. The literature about parallel graph coloring refers mainly to the algorithm using barrier synchronization. However, they do not prove the correctness of the algorithm. This is crucial because all the threads run in an asynchronous manner and the behaviour of a thread at a particular time instant cannot be determined. In this paper, we present a modified version of this algorithm in section 3 and give a proof of correctness of the algorithm in section 4.\\
\indent Subsequently, we put forth a new approach to deal with this problem using some standard locking techniques in section 3. We then evaluate the performance of the proposed algorithm against the barrier synchronization algorithm as well as the sequential coloring algorithm in section 5. We see that the performance of our proposed approach outweighs the other algorithms.
\section{Background}
\subsection{Problem Definition}
A graph $G$ is a pair $(V, E)$ of a set of vertices $V$ and a set of edges $E$. The edges are unordered pairs of the form $\{i, j\}$ where $i, j$ $\in$ $V$. Two vertices $i$ and $j$ are said to be adjacent if and only if $\{i, j\}$ $\in$ $E$ and non-adjacent otherwise.\\
The degree of a vertex $v$ is the number of vertices adjacent to $v$ and is denoted by deg($v$). The maximum and minimum degree in a graph $G$ are denoted by $\Delta$ and $\delta$ respectively. An independent set in a graph is a set of vertices that are mutually non-adjacent. This means that there is no edge between any pair of vertices in an independent set.\\
\newline
\textbf{The Graph Coloring Problem:}\\
A vertex coloring of a simple graph, or simply coloring for short, is an assignment of colors to the vertices such that no two adjacent vertices are assigned the same color. Alternatively, a coloring is a partition of the vertex set into a collection of vertex-disjoint independent sets. Each independent set in such a partition is called a color class. The graph coloring problem is then to find a vertex coloring for a simple graph using the minimum number of colors possible. It is easy to see that any arbitrary simple graph can be colored with $\Delta$ + 1 colors.
\subsection{Related Work}
The problem of parallel graph coloring has been studied extensively. Jones \& Plassman \cite{DBLP:journals/siamsc/JonesP93} process the vertices of a graph in a random order. The difficulty with this approach lies in identifying the most effective ordering of the vertices according to the graph in question. Gebremedhin, Manne \& Woods  \cite{gebremedhin2006speeding} describe an algorithm for parallel graph coloring by the technique of conflict detection. The major drawback here lies in the fact that all the vertices in color conflict are colored sequentially. {\c{C}}ataly{\"{u}}rek et al. \cite{DBLP:journals/pc/CatalyurekFGHP12} extend the previous notion and increase the parallelism in the algorithm. However, they do not provide a proof as to why the algorithm would terminate in a finite number of steps or would result in proper coloring of the graph eventually. Also, the paper does not compare the speedup of the proposed algorithm against the sequential coloring to see if the goal in parallelizing the algorithm is achieved.
\subsection{System Model}
The RAM (Random Access Machine) is a model used successfully to predict the performance of programs on single processor (sequential) computers. A natural extension of this model for parallel computers is the shared-memory model. This model consists of a number of processors/threads, each of which has its own local memory and can execute its own local program, and all of which communicate by exchanging data through a shared memory unit, also called global memory. In this paper, we assume that our system consists of $n$ processors, accessed by $p$ threads/processors that run in a completely asynchronous manner. Hence, we make no assumption about the relative speeds of the processors. We also assume that none of these processors and threads fail.

\section{Solution Approach}
In this section, we present various approaches suitable for dealing with this problem on a shared programming memory model. In the first approach, we extend the notion of the algorithm in \cite{DBLP:journals/pc/CatalyurekFGHP12} and also prove the correctness in the later section. The major difference in our algorithm is that \cite{DBLP:journals/pc/CatalyurekFGHP12} has not explicitly used barriers for synchronization. We use barriers because it is more efficient than creating new threads as in \cite{DBLP:journals/pc/CatalyurekFGHP12}. Later, we propose a new algorithm using the standard locking techniques and compare it against the first one. In the following algorithms, we have used the First Fit Coloring strategy. This algorithm assigns each vertex the least legal color.
\subsection{Using Barrier Synchronisation}
\indent Let us say that the vertex ids are from \{1, 2, \dots, $\mid$$V$$\mid$\}. Initially, the graph $G$ = ($V,E$) is preprocessed by partitioning it uniformly into $p$ blocks where $p$ is the number of threads. Then vertices in \{1, \dots, $|V| / p$\} get assigned to $V_1$, \{$|V| / p+1$, \dots, 2$|V| / p$\} get assigned to $V_2$ and so on until $V_p$. This method of partitioning helps in simplifying the proof of correctness. The vertices in each partition/block are classified into:- internal vertices (whose all the neighbouring vertices lie in the same partition) and boundary vertices (who have  neighbours belonging to other partitions). Each thread is responsible for proper coloring of vertices in its partition. \\
\indent The algorithm has two phases: tentative coloring \& conflict detection phase. Each thread maintains a copy of colors assigned to its neighbours locally in \textit{ForbiddenColors} List. In the first phase, a thread takes into account all its previously colored neighbours from the local copy to assign a color. However, it might result in two threads simultaneously coloring the vertices adjacent to each other with same color. Hence, in the second phase, each thread $T_i$ checks whether the vertices $V_i$ are assigned valid colors by comparing the color of a vertex against all its neighbours that were colored in first phase. If any vertex and its neighbor have the same color, then the vertex in lower partition is recolored.\\
\indent The first and second phase are synchronized by a barrier that ensures that all the $p$ threads start their execution at the same instant. This is crucial because if a thread were still coloring while other tries to detect conflicts, then this can lead to false detection eventually leading to improper coloring.\\
\newline
Algorithm \ref{alg1} describes the pseduo-code of the modified barrier synchronization algorithm.\\

\begin{algorithm}[H]
%\begin{multicols}{2}

\caption{Using Barrier}
\label{alg1}
\begin{algorithmic}[1]

\State \textbf{Input:} $p$ $\leftarrow$ no of threads
\State uniform partitioning of $V$ into $V_1$, $V_2$, \dots, $V_p$ in increasing order of vertex ids
\State $m$ $\leftarrow$ maximum degree of graph
\Procedure{ParallelGraphColoring}{$G$ = ($V,E$)}
\For{all thread $T_i$ $\mid$ $i$ $\in$ \{1, ..., $p$\}}
\State Identify boundary vertices of partition $i$
\State Initialise \textit{TotalColors}[$m$ + 1] = \{0, 1, \dots, $m$\}
\For{$v$ $\in$ $V_i$}
\State Create List $v$.\textit{ForbiddenColors} of size $=$ adjacent($v$)
\State Initialise $v$.\textit{ForbiddenColors} to $-1$ 
\EndFor
\State $U_i$ $\leftarrow$ $V_i$
\While{$U_i$ $\neq$ $\emptyset$}
\For{each $v$ $\in$ $U_i$} \Comment{Phase 1 starts}
\State color($v$) $\leftarrow$ min\{\textit{TotalColors} $-$ $v$.\textit{ForbiddenColors}\}
\For{each $u$ $\in$ adjacent($v$) where $u$ $\in$ $V_i$}
\State Update color($v$) in $u$.\textit{ForbiddenColors}
\EndFor
\EndFor
\State Wait for all threads to reach here 
\Comment{Using barrier}
\algstore{myalg}
\end{algorithmic}
\end{algorithm}

\begin{algorithm}                     
\begin{algorithmic} [1]                   % enter the algorithmic environment
\algrestore{myalg}

\State $R_i$ $\leftarrow$ $\emptyset$ 
\Comment{Phase 2 starts}

\For{each $v$ $\in$ $U_i$ $\mid$ $v$ is a boundary vertex in $U_i$}
\For{each $u$ $\in$ adjacent($v$) where $u$ $\notin$ $V_i$}
\State Update color($u$) in $v$.\textit{ForbiddenColors}
\If {color($u$) = color($v$) $\mid$ $u$ $\in$ $V_j$ and $i$ $<$ $j$}
\State $R_i$ $\leftarrow$ $R_i$ $\cup$ \{$u$\}
\EndIf
\EndFor
\EndFor
\State $U_i$ $\leftarrow$ $R_i$
\State Wait for all threads to reach here 
\Comment{Using barrier}
\EndWhile
\EndFor
\EndProcedure
\end{algorithmic}
%\end{multicols}

\end{algorithm}

\subsection{Using Locks}
The motivation behind using an alternative to the barrier synchronization approach lies in the fact that, since all the threads get synchronized at two points in each iteration (line 20 and line 31 in Algorithm \ref{alg1}), there is a major unfavourable impact on the performance of the algorithm. The overall goal is to avoid global synchronization of threads and let them run independently. We present an algorithm based on the locking of the graph vertices. With locks, coloring the vertex becomes a critical section and a thread can only enter the critical section when it acquires the lock.\\
This can be achieved in two ways:- Coarse and Fine Grained Locking. 

\subsubsection{Coarse Grained Locking}
With Coarse Grained Locking, there exists a big lock on the complete list of boundary vertices. This implies that at any point, a vertex must acquire a lock on this list to color itself. The pseudo code of the algorithm is described below:

\begin{algorithm}
\caption{Using Coarse Grained Locks}
\label{alg2}

\begin{algorithmic}[1]
\State \textbf{Input:} $p$ $\leftarrow$ no of threads
\State uniform random partitioning of $V$ into $V_1$, $V_2$, \dots, $V_p$
\State $m$ $\leftarrow$ maximum degree of graph
\State List $B$ $\leftarrow$ all boundary vertices of graph $G$
\Procedure{ParallelGraphColoring}{$G$ = ($V,E$)}
\For{all thread $T_i$ $\mid$ $i$ $\in$ \{1, \dots, $p$\}}
\State Initialise \textit{TotalColors}[$m$ + 1] $=$ \{0, 1, \dots, $m$\}
\For{each $v$ $\in$ $V_i$ $\mid$ $v$ is a internal vertex in $V_i$}
\State color($v$) $\leftarrow$ min\{\textit{TotalColors} $-$ color(adjacent($v$)\}
%\State \parbox[t]{\dimexpr\linewidth-\algorithmicindent}{Assign $v$ a least color from \textit{TotalColors} different from color(adj($v$))\strut}
\EndFor
\For{each $v$ $\in$ $V_i$ $\mid$ $v$ is a boundary vertex in $V_i$}
\State \textit{Lock} List B
\State color($v$) $\leftarrow$ min\{\textit{TotalColors} $-$ color(adjacent($v$)\}
%\State \parbox[t]{\dimexpr\linewidth-\algorithmicindent}{Assign $v$ a least color from \textit{TotalColors} different from color(adj($v$))\strut}
\State \textit{Unlock} List B
\EndFor
\EndFor
\EndProcedure
\end{algorithmic}
\end{algorithm}

\subsubsection{Fine Grained Locking}
Coarse Grained Locking can be improvised on by making use of fine grained locks wherein each vertex has a lock. A thread wishing to color the neighboring vertices of a vertex, has to obtain the corresponding locks. However, to avoid deadlock, a global ordering of vertices is maintained (based on their vertex ids) and vertices acquire locks in the respective order. 
The pseudo code of the algorithm is described below:

\begin{algorithm}
\caption{Using Fine Grained Locks}
\label{alg3}

\begin{algorithmic}[1]
\State \textbf{Input:} $p$ $\leftarrow$ no of threads
\State uniform random partitioning of $V$ in $V_1$, $V_2$, . . ,$V_p$
\State $m$ $\leftarrow$ maximum degree of graph
\Comment{Vertices are inherently ordered by their vertex ids'}
\Procedure{ParallelGraphColoring}{$G$ = ($V,E$)}
\For{all thread $T_i$ $\mid$ $i$ $\in$ \{1,\dots,$p$\}}
\State Identify boundary vertices in $V_i$
\State Initialise \textit{TotalColors}[$m$ + 1] = \{0, 1, ...., m\}
\For{each $v$ $\in$ $V_i$ $\mid$ $v$ is a internal vertex in $V_i$}
\State color($v$) $\leftarrow$ min\{\textit{TotalColors} $-$ color(adjacent($v$)\}
%\State \parbox[t]{\dimexpr\linewidth-\algorithmicindent}{Assign $v$ a least color from \textit{TotalColors} different from color(adj($v$))\strut}
\EndFor
\For{each $v$ $\in$ $V_i$ $\mid$ $v$ is a boundary vertex in $V_i$}
\State List $A_i$ $\leftarrow$ adj($v$)
%\State Sort A on vertex id
\State $A_i$ $\leftarrow$ $A_i$ $\cup$ \{$v$\}
\State \textit{Lock} all vertices in $A_i$ in increasing order of vertex ids
\State color($v$) $\leftarrow$ min\{\textit{TotalColors} $-$ color(adjacent($v$)\}
%\State \parbox[t]{\dimexpr\linewidth-\algorithmicindent}{Assign $v$ a least color from \textit{TotalColors} different from color(adj($v$))\strut}
\State \textit{Unlock} all vertices in $A_i$
\EndFor
\EndFor
\EndProcedure
\end{algorithmic}
\end{algorithm}
\vspace{-0.5cm}
\section{Proof of Correctness}
\textbf{Lemma 1:} \textit{Barrier Algorithm results in proper coloring of the graph.}\\\\
\textbf{Proof:} Let us prove by contradiction. So we assume that the algorithm did not result in proper coloring meaning that two adjacent vertices of the graph have the same color. Each round/iteration of the algorithm consist of 2 phases: coloring and conflict detection phases respectively. We denote the coloring phase of $i^{th}$ iteration as $i.1$ and conflict detection phase of $i^{th}$ iteration as $i.2$.\\
\newline
Let us say that a vertex $v_x$ gets colored $c$ in round $i.1$ and vertex $v_y$ gets colored the same in round $j.1$, both belonging to different partitions and i $\leq$ j.\\
\begin{center}
$color^{i.1}$($v_x$) = $color^{j.1}$($v_y$) = $c$ where $v_x$, $v_y$ belong to different partitions
\end{center}
Now there are two possibilities as follows:\\
\newline
a) $v_x$ was assigned color $c$ in round $(j-1).1$: In this case, in round $(j-1).2$, $v_x$ and $v_y$ would be identified with same color and the vertex in the lower partition id would get recolored in round $j.1$. Hence either $v_x$ or $v_y$ would have a color different from $c$. Also since $i$ $\leq$ $j$, this means that both vertices get properly colored. Hence this is a contradiction to our initial assumption.\\
\newline
b) $v_x$ was assigned a color different from $c$ in round $(j-1).1$: In this case, $v_x$ got recolored back to color $c$ in round $j$.1, then a conflict will be detected in round $j$.2 and it will be resolved in round $(j+1)$.1. Hence it again contradicts our assumption.\\
\\Thus we can conclude that eventually all the conflicts get resolved and no two adjacent vertices get assigned to a same color.\\
\newline
\textbf{Lemma 2:} \textit{Barrier Algorithm terminates after a maximum of $p+1$ iterations.}\\\\
\textbf{Proof:} The partitions of the graph are $V_1$, $V_2$, ... , $V_p$. In each round, vertex in $V_i$ is recolored if it has a conflict with a vertex in $V_{i+1}$, ... , $V_p$.\\
\newline
In the $1^{st}$ round, at least $V_p$ gets properly colored and all conflicts in $V_{p-1}$ with $V_p$ are identified, which are resolved in the next round. \\Similarly, in the $2^{nd}$ round, $V_{p-1}$ gets properly colored and all conflicts in $V_{p-2}$ with $V_{p-1}$ are identified, which are resolved in the next round. \\
\newline
Thus it is easy to see that after $p+1$ iterations, $V_1$ gets properly colored. Also, the maximum number of times a vertex in partition $V_i$ gets recolored is ($p - i$). \\
\newline
\textbf{Corollary:} \textit{Number of iterations of the barrier algorithm is upper bounded by the number of partitions of the graph.}

\section{Simulation Results}
\subsection{Experimental Setup}
For testing the performance, we have considered 24 core Intel Xeon (X5675) running at 3.07 GHz core frequency. Each core supports 6 hardware threads. While considering the time taken for coloring the graph in the multi-threaded version, we have included the time taken for partitioning of graph as well. However, time taken for coloring in all versions (sequential \& parallel) excludes the time taken to read the graph input. Each data point is obtained after averaging for 10 iterations.
\subsection{Performance Graphs}
To test the performance of the algorithms, we have used datasets of real world graphs from \cite{snapnets}. We have considered two dataset belonging to real world:- Live Journal and Orkut Community from SNAP. We have evaluated them against two metrics: Time Taken to color the graph and Number of Colors Used. We have tested each of the dataset against 4 versions of algorithms: sequential run, barrier synchronization algorithm, coarse grained locking and fine grained locking.\\
On the x-axis, we have varied the number of threads from 1 to 1000. For sequential run, we have executed it for 1 thread but represented it as a line, for reference against the other algorithms. The single data label of each color represents the best data value that the respective algorithm performs. \\

\begin{center}
\begin{figure}[H]
 \epsfig{width=16cm,height=7cm,figure=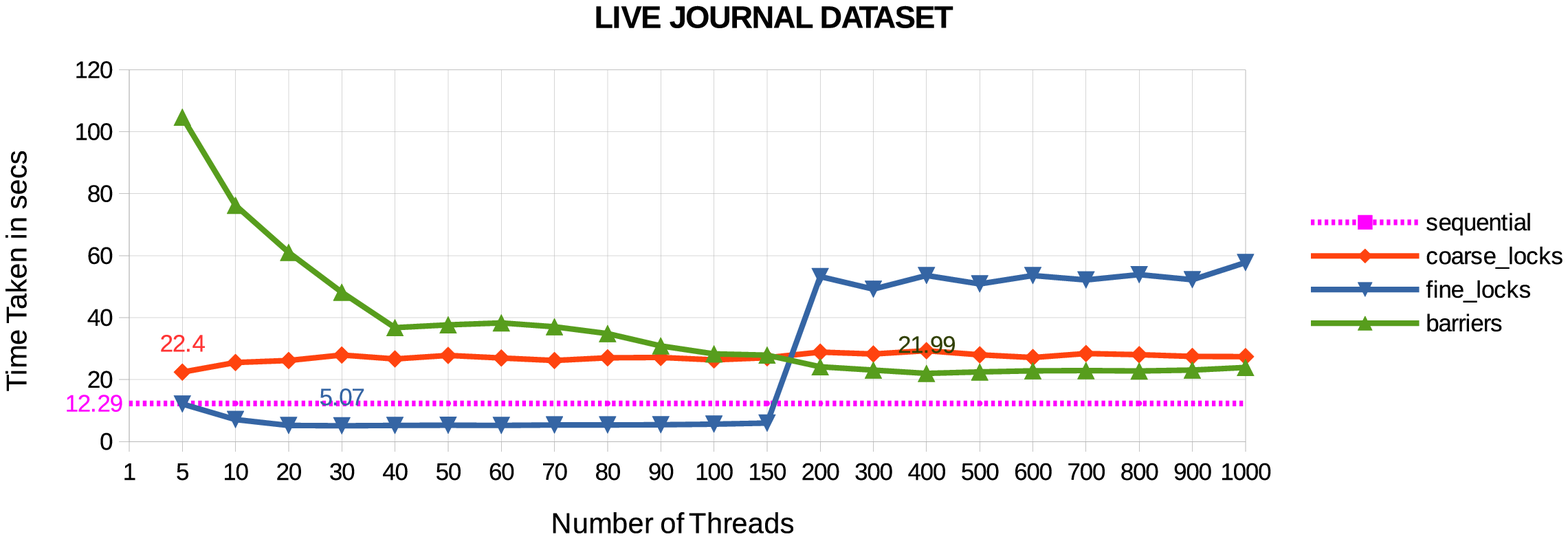}
 \vspace{-0.9cm}
 \centering
 \caption{Time Taken in secs (with barriers) v/s Number of Threads}
\label{live1}
 \end{figure}
\end{center}
\vspace{-1.5cm}
\begin{center}
\begin{figure}[H]
 \epsfig{width=16cm,height=7cm,figure=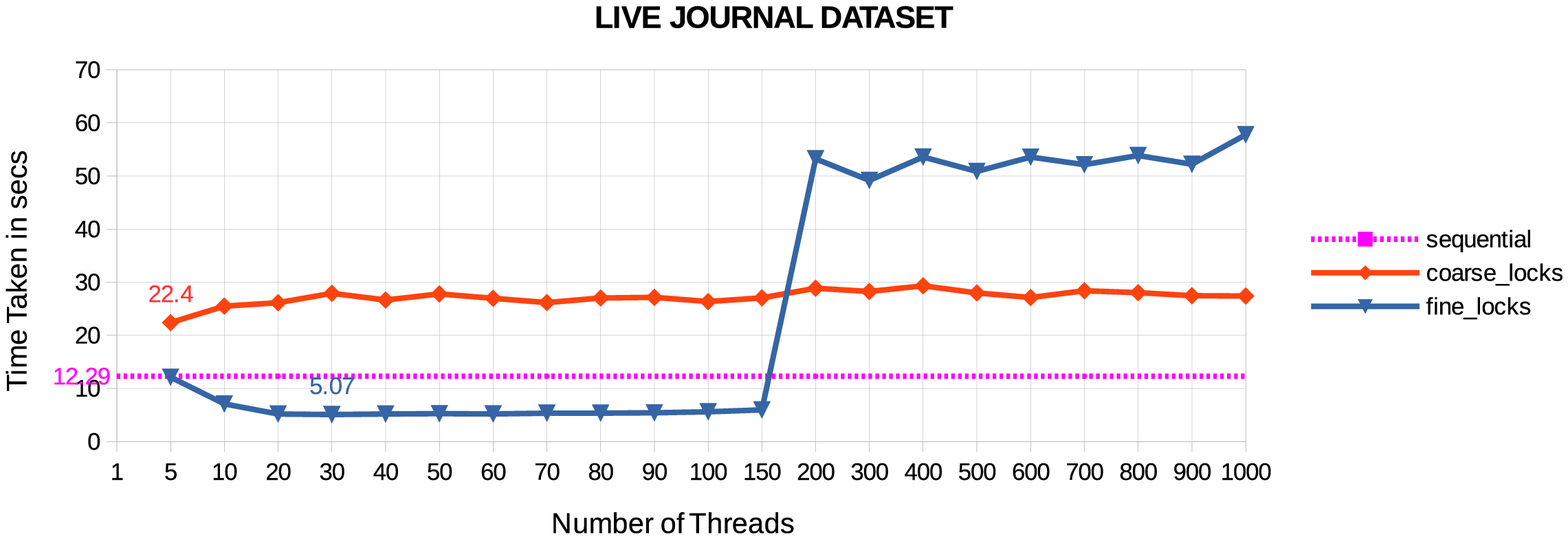}
 \vspace{-0.9cm}
 \caption{Time Taken in secs (without barriers) v/s Number of Threads}
\label{live2}
 \end{figure}
\end{center}
\vspace{-1.4cm}
\begin{center}
\begin{figure}[H]
 \epsfig{width=16cm,height=7cm,figure=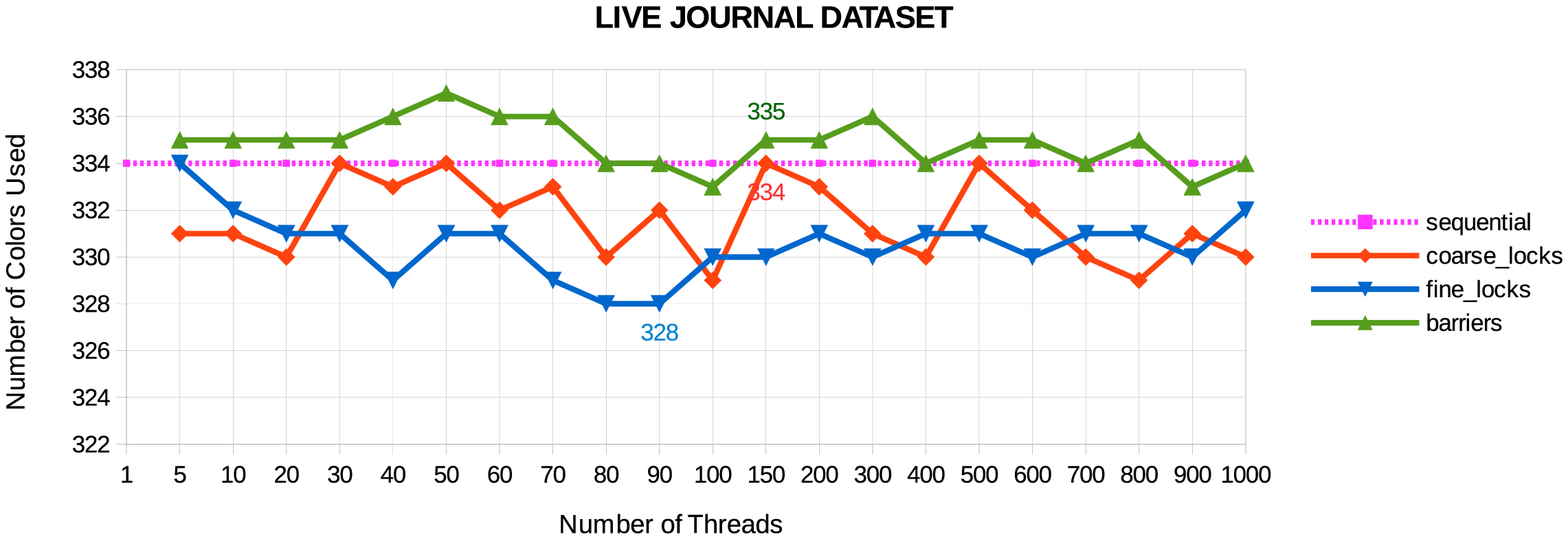}
 \vspace{-0.9cm}
 \caption{Number of Colors Used v/s Number of Threads}
\label{live3}
 \end{figure}
\end{center}
\vspace{-1.2cm}
\begin{center}
\begin{figure}[H]
 \epsfig{width=16cm,height=6.5cm,figure=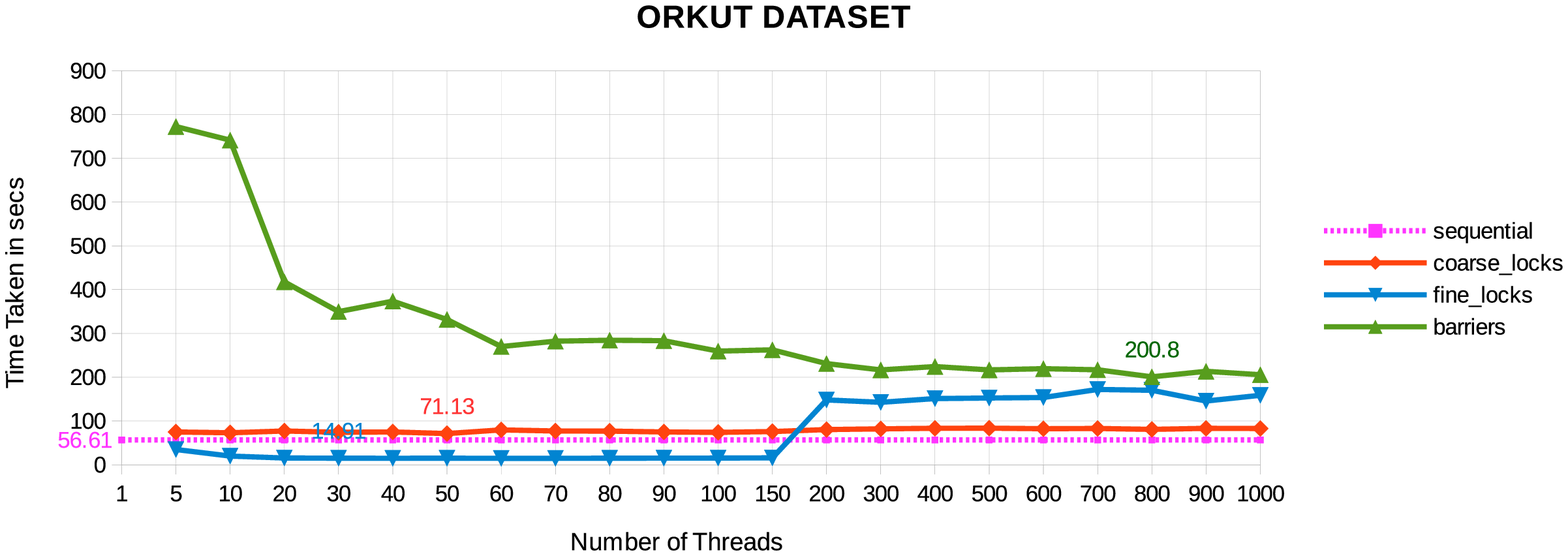}
 \vspace{-0.9cm}
 \caption{Time Taken in secs (with barriers) v/s Number of Threads}
\label{orkut1}
 \end{figure}
\end{center}
\vspace{-1.5cm}
\begin{center}
\begin{figure}[H]
 \epsfig{width=16cm,height=6.5cm,figure=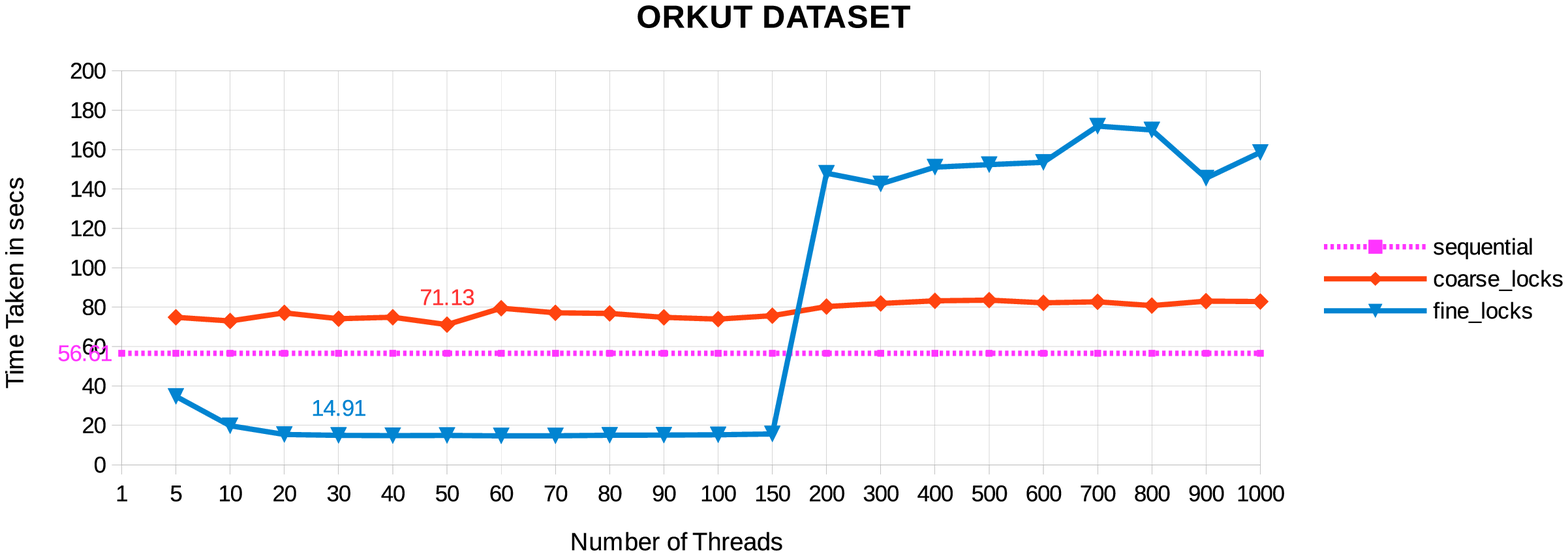}
 \vspace{-0.9cm}
 \caption{Time Taken in secs (without barriers) v/s Number of Threads}
\label{orkut2}
 \end{figure}
\end{center}
\vspace{-1.5cm}
\begin{center}
\begin{figure}[H]
 \epsfig{width=16cm,height=6.5cm,figure=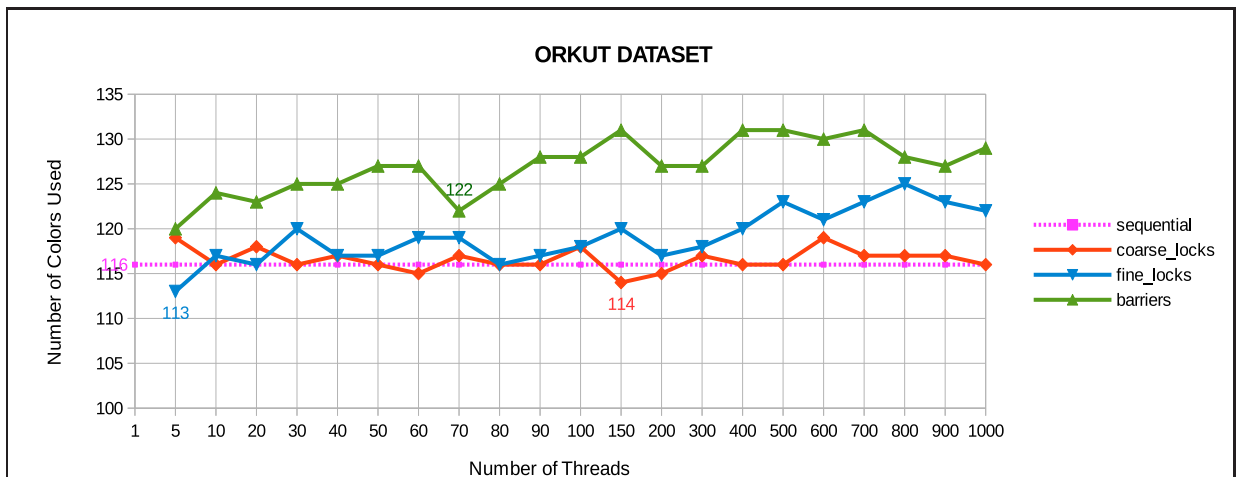}
 \vspace{-0.9cm}
 \caption{Number of Colors Used v/s Number of Threads}
\label{orkut3}
 \end{figure}
\end{center}
\vspace{-1cm}
\subsection{Result Analysis}
As can be clearly observed from the performance graphs in the previous subsection, the barrier synchronization approach does not fare well and is not comparable to the sequential coloring. On the other hand, locks seem to perform fairly well. We observe that fine grained locking performs significantly better as compared to the sequential coloring both in terms of time taken as well as the number of colors used.\\
\indent Speedup of a parallel algorithm is defined as the ratio of time taken by the sequential run to time taken by the parallel algorithm. We see that speedup for barrier algorithm is 0.56 and 0.28 for live journal and orkut datasets. Whereas, the speedup obtained from locking algorithm is 2.42 and 3.79 respectively.\\
\indent An interesting point to note is that the performance of the fine grained locking algorithm deteriorates after the number of threads exceed 150. One possible reason to explain it is that the hardware architecture supports only 24 * 6 $\approx$ 150 (roughly) hardware threads. Another reason could be there are long waiting chains formed due to locking. 
\vspace{-.2cm}
\section{Conclusion \& Future Work}
We have presented parallel algorithms for graph coloring suitable to shared memory programming model. We have looked into the most commonly used approach for coloring using barrier synchronization and also given the proof of correctness of the same. We have also proposed a new approach using locks. Using the SNAP dataset, we evaluated the performance of the algorithms on the Intel platform. The results show that the improvement is noteworthy. This gives a motivation that the overhead of locking and unlocking operations is very less as compared to the overhead due to barrier synchronization.\\
\indent In this paper, we have used the First Fit Coloring strategy. We intend to extend the algorithms with other coloring schemes to reduce and tightly upper bound the number of colors used. We also aim to test with different kinds of datasets comprising of dense graphs. Also, we plan to improve fine grained locking by resolving the issue of long waiting chains which leads to the deterioration in the performance of the algorithm after the number of threads exceed 150.
\vspace{-.3cm}
\bibliographystyle{plain}
\bibliography{biblio}

\begin{thebibliography}{1}

\bibitem{boman2005scalable}
Erik~G Boman, Doruk Bozda{\u{g}}, Umit Catalyurek, Assefaw~H Gebremedhin, and
  Fredrik Manne.
\newblock A scalable parallel graph coloring algorithm for distributed memory
  computers.
\newblock In {\em Euro-Par 2005 Parallel Processing}, pages 241--251. Springer,
  2005.

\bibitem{DBLP:journals/pc/CatalyurekFGHP12}
{\"{U}}mit~V. {\c{C}}ataly{\"{u}}rek, John Feo, Assefaw~Hadish Gebremedhin,
  Mahantesh Halappanavar, and Alex Pothen.
\newblock Graph coloring algorithms for multi-core and massively multithreaded
  architectures.
\newblock {\em Parallel Computing}, 38(10-11):576--594, 2012.

\bibitem{gebremedhin2006speeding}
Assefaw~H Gebremedhin, Fredrik Manne, and Tom Woods.
\newblock Speeding up parallel graph coloring.
\newblock In {\em Applied Parallel Computing. State of the Art in Scientific
  Computing}, pages 1079--1088. Springer, 2006.

\bibitem{DBLP:journals/concurrency/GebremedhinM00}
Assefaw~Hadish Gebremedhin and Fredrik Manne.
\newblock Scalable parallel graph coloring algorithms.
\newblock {\em Concurrency - Practice and Experience}, 12(12):1131--1146, 2000.

\bibitem{DBLP:journals/siamsc/JonesP93}
Mark~T. Jones and Paul~E. Plassmann.
\newblock A parallel graph coloring heuristic.
\newblock {\em {SIAM} J. Scientific Computing}, 14(3):654--669, 1993.

\bibitem{snapnets}
Jure Leskovec and Andrej Krevl.
\newblock {SNAP Datasets}: {Stanford} large network dataset collection.
      \newblock \url{http://snap.stanford.edu/data}, June 2014.

\bibitem{patwary2011new}
Md~Mostofa~Ali Patwary, Assefaw~H Gebremedhin, and Alex Pothen.
\newblock New multithreaded ordering and coloring algorithms for multicore
  architectures.
\newblock In {\em Euro-Par 2011 Parallel Processing}, pages 250--262. Springer,
  2011.

\end{thebibliography}
%\end{multicols}

\end{document}